\documentclass{elsart}
\usepackage{harvard}
\usepackage{graphicx}

\usepackage{amssymb}

\def\url#1{{\ttfamily\def\/{/\discretionary{}{}{}}#1}}

\begin{document}
	
\begin{frontmatter}
\title{Multi-Resolution Imaging and Spectra
of Extended Sources}

\author[Rudnick]{Lawrence Rudnick\thanksref{lr}}

\thanks[lr]{E-mail:  larry@tc.umn.edu}

\address[Rudnick]{Department of Astronomy, University of Minnesota, 116 
Church St. SE, Minneapolis, MN 55455-0149}

\begin{abstract}
I introduce a straightforward technique for the filtering of extended astronomical 
images into components of different spatial scales.  For a positive
original image, each component
is positive definite, and the sum of all components equals the
original image.  In this way, the components are each individually suitable for
flux measurements and broadband spectra calculations.  I present an illustration
of this technique on the radio galaxy Cygnus~A.
\end{abstract}
\end{frontmatter}

\section{Introduction}
\label{intro}
 One common problem in astronomical image
processing is the detection of 
point or small-scale features against a varying background, or more 
generally, the separation of information on different spatial scales.
I introduce a technique for filtering extended astronomical images based 
on a suggestion in \citeasnoun{sta98}.  Images can be decomposed into an arbitrary 
number of components, each representing structures on different spatial 
scales and whose sum equals the original image.  Each component is positive definite and 
therefore can be suitable, e.g.,  for spectral index or optical color studies.
The method is similar in some ways to the multi-scale wavelet analysis
discussed by \cite{slezak}.

I have applied the multiresolution filtering 
to a number of radio galaxies.  I find that 
the spectra of fine-scale features do not follow a simple pattern, as 
opposed to the slow gradients which often characterize the more diffuse 
emission.  In this paper, I will show some results on Cygnus~A as an 
example.

\section{The Method}

For the simplest two-component image separation, the initial image is 
decomposed into {\it Filtered} and {\it Open} maps, where $Original = Filtered + Open$.

The {\it Open} map is made by first replacing each pixel on the original map 
by the minimum value in a box of fixed size which slides across the map, 
(a process called 
``erosion"), followed by replacing each pixel on the ``eroded" map by the 
maximum value in a sliding box across the map.  The result is the {\it Open}
map.  Then, $Filtered = Original - Open$.

The {\it Filtered} map now contains only structures with sizes smaller than or 
on the order of the sliding box size.  Everything else, by construction, is 
in the {\it Open} map.  The method can be generalized by decomposing the 
{\it Open} map itself into an additional {\it Filtered'} and {\it 
Open'} map by operating with a 
larger box on the {\it Open} map.  This method can 
be generalized to an arbitrary number of steps.  In practice, I have 
found that a change in box size of approximately four between steps is
necessary to isolate independent components.

This technique has been implemented in AIPS with a trivial 
modification to the task MWFLT and a RUN file to automate the various
steps.  It would also be quite straightforward to implement, e.g., in 
IDL.  Readers are encouraged to contact the author for details.


\section{The Use of Multiresolution Filtering for Spectra}
The measurement of spectra in extragalactic radio sources is fraught 
with problems such as the confusion between overlapping structures, 
biases due to bowls, and the subjectivity of modelling or other 
techniques used to isolate features.  Each spectral measurement 
technique, e.g.,  direct division, the regression, or ``T-T'' method,
or even the tomography we have introduced \cite{kat97}, has its own 
particular problems.  Multiresolution filtering offers a complementary
method which can be useful when applied with care.  I have done a 
great deal of testing of the robustness of this technique, as will be 
described in an upcoming paper.  An illustration of our results on
Cygnus~A is shown in Figure 1.

To calculate the spectral index, I first decompose the maps into
two or more components.  Then, the spectral index can be calculated
separately for each component, either by straight division or by
regression.  In cases where there is a clear bimodal (or more)
distribution of spatial scales, this is a meaningful process.  If,
however, there is a continuous range of spatial scales, then the
decomposition and the resultant spectra are somewhat arbitrary.  
They still may be useful, however, for seeing whether there are
differences between fine- and large- scale spectra.

In general, I find that fine-scale structure has a slightly flatter
spectral index than the local large-scale structure. For Cygnus~A,
however, there are locations in the source where the opposite is
true.  The one-dimensional plots at the bottom of Figure 1 do not
show the full range of spectra, but still give an indication of 
the richness of the fine-scale spectral structure. 

Looking at the spectra of the large scale emission (which might be
most appropriate for ageing analyses if we could eliminate the
problems due to magnetic field variations), we can use this
filtering technique to assess/eliminate problems with confusion from
small scale structures.  In the case of Cygnus~A, this turns out not
to be a significant problem, as seen by the close correspondence between
the spectra of the original (not decomposed) and {\it open} maps.

 How can the large-scale and fine-scale
spectra be different?  First, as seen in the numerical simulations
such as those presented here by Jones, Tregillis and Ryu, there can
be filamentary or other structures with very different particle
acceleration and/or loss histories than neighboring features.
Alternatively, the local magnetic fields may vary between structures,
causing a change in the observed spectral index even if the
relativistic particle populations are identical.  It is this confusion
which causes major problems for standard ageing analyses.

I have also worked with colleagues to experiment with galactic plane
continuum data and deep optical/IR images.  It appears that multiresolution
filtering will be useful for a wide range of astronomical images.

This work is supported by NSF grant AST 96-16964 and AST AST 96-19438
at the University of Minnesota.  I am grateful to John Dickey, Andrew
Young, Barron Koralesky and Deb Katz-Stone for many useful ideas and
feedback.

\begin{figure} 
\begin{center} 
\rotatebox{0}{\scalebox{.7}[.7]{\includegraphics{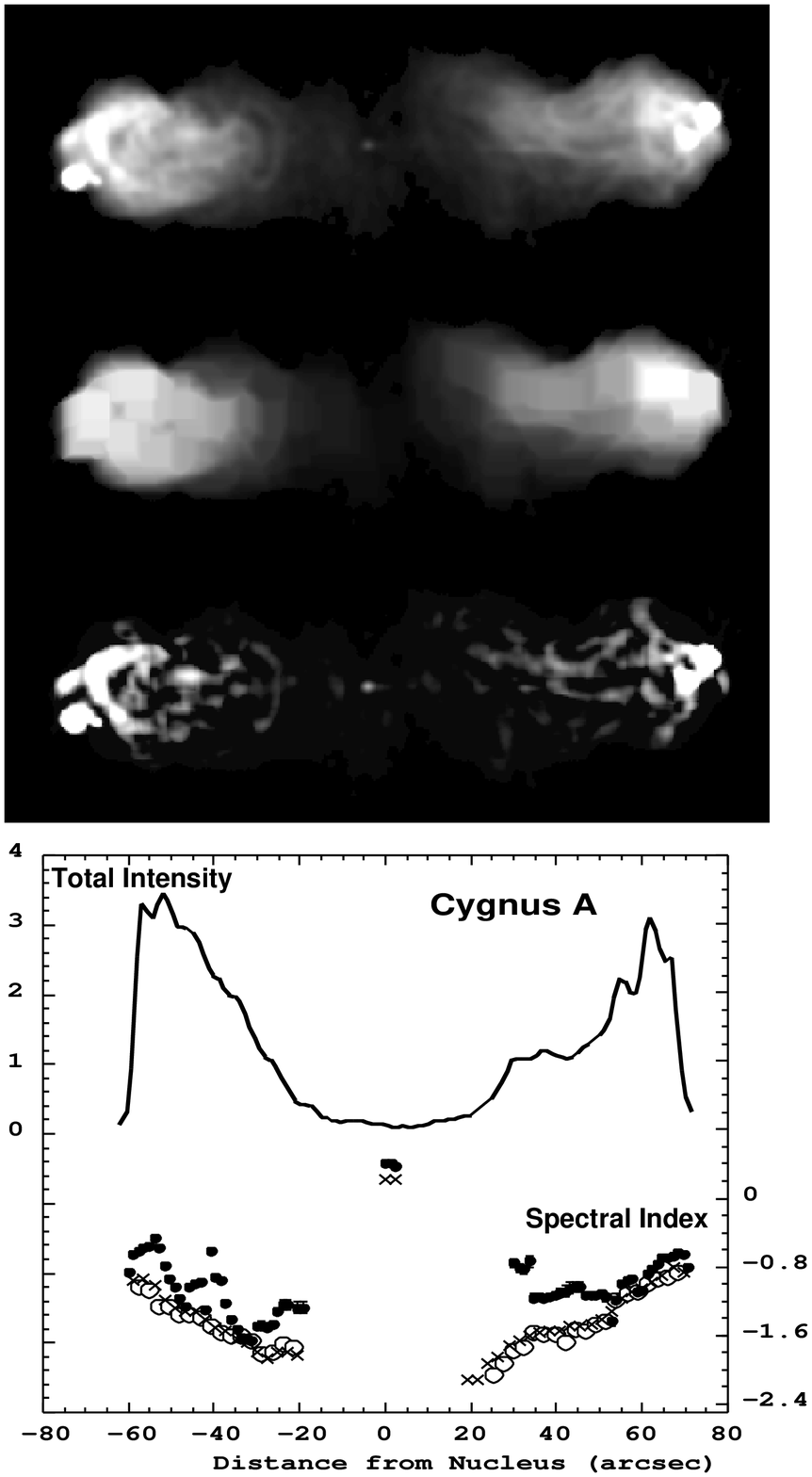}}}
\caption{Filtering results on Cygnus~A. (All based on data from Carilli
et al., 1991.) The top image is the total intensity at $\lambda 20cm$
with a resolution of 1.5''.  The middle image is the  {\it open} map and the
bottom the {\it filtered} map with a box size of 4.8''.  The plots at
the bottom show the total intensity in transverse strips along the
major axis along with three different measures of the spectral index
between $\lambda 20cm$ and $\lambda 6cm$.  The x's show the indices from
the original maps at two freqeuncies; the open circles from the corresponding {\it
open} maps, and the filled circles from the {\it filtered} maps.}
\label{cygmulti} \end{center} \end{figure}


\end{document}